\documentclass[final,3p,times,onecolumn]{elsarticle}

\usepackage{natbib,twoopt}
\usepackage{graphicx}

\usepackage{amsmath}
\usepackage{amssymb}

\usepackage{textcomp}

\usepackage{dcolumn}
\usepackage{bm}

\usepackage[english]{babel}
\usepackage{verbatim} 

\usepackage{titlesec}

\usepackage{color}
\usepackage{longtable}

\usepackage{latexsym}

\usepackage{eqnarray,amsmath}

\usepackage{mathtools}
\usepackage{empheq}

\usepackage{tikz-cd}

\journal{Elsevier}

\begin{document}

\begin{frontmatter}

\title{The effect of the junction model on the anomalous diffusion\\ 
in the 3D comb structure}

\author{A.R. Dzhanoev}
\address{Universit\"at Potsdam, Karl-Liebknecht-Str.24/25, 14476 Potsdam-Golm, Germany }

\author{I.M. Sokolov}
\address{Institut f\"ur Physik, Humboldt-Universit\"at zu Berlin, Newtonstra{\ss}e 15, D-12489 Berlin, Germany}


\begin{abstract}
The diffusion in the comb structures is a popular model of geometrically induced anomalous diffusion. 
In the present work we concentrate on the 
diffusion along the backbone in a system where sidebranches are planes, and the diffusion thereon is anomalous and 
described by continuous time random walks (CTRW). 
We show that the mean squared displacement (MSD) in the backbone of the comb behaves 
differently depending on whether the waiting time periods in the sidebranches are reset after the step in the backbone is done 
(a rejuvenating junction model), or not (a non-rejuvenating junction model). In the rejuvenating case the subdiffusion in the 
sidebranches only changes the prefactor in the ultra-slow (logarithmic) diffusion along the backbone, while in the 
non-rejuvenating case the ultraslow, logarithmic subdiffusion is changed to a much faster power-law subdiffusion 
(with a logarithmic correction) as it was found earlier by Iomin and Mendez [Chaos Solitons and Fractals 2016; 82:142]. 
Moreover, in the first case the result does not change if the 
diffusion in the backbone is itself anomalous, while in the second case it does. 
Two of the special cases of the considered models (the non-rejuvenating junction under normal diffusion in the backbone, and
rejuvenating junction for the same waiting time distribution in the sidebranches and in junction points) 
were also investigated within the approach based on the corresponding generalized Fokker-Planck equations. 
\end{abstract}


\end{frontmatter}

\section{Introduction}
\label{}

\begin{figure}[htbp!]
\includegraphics[scale=0.35]{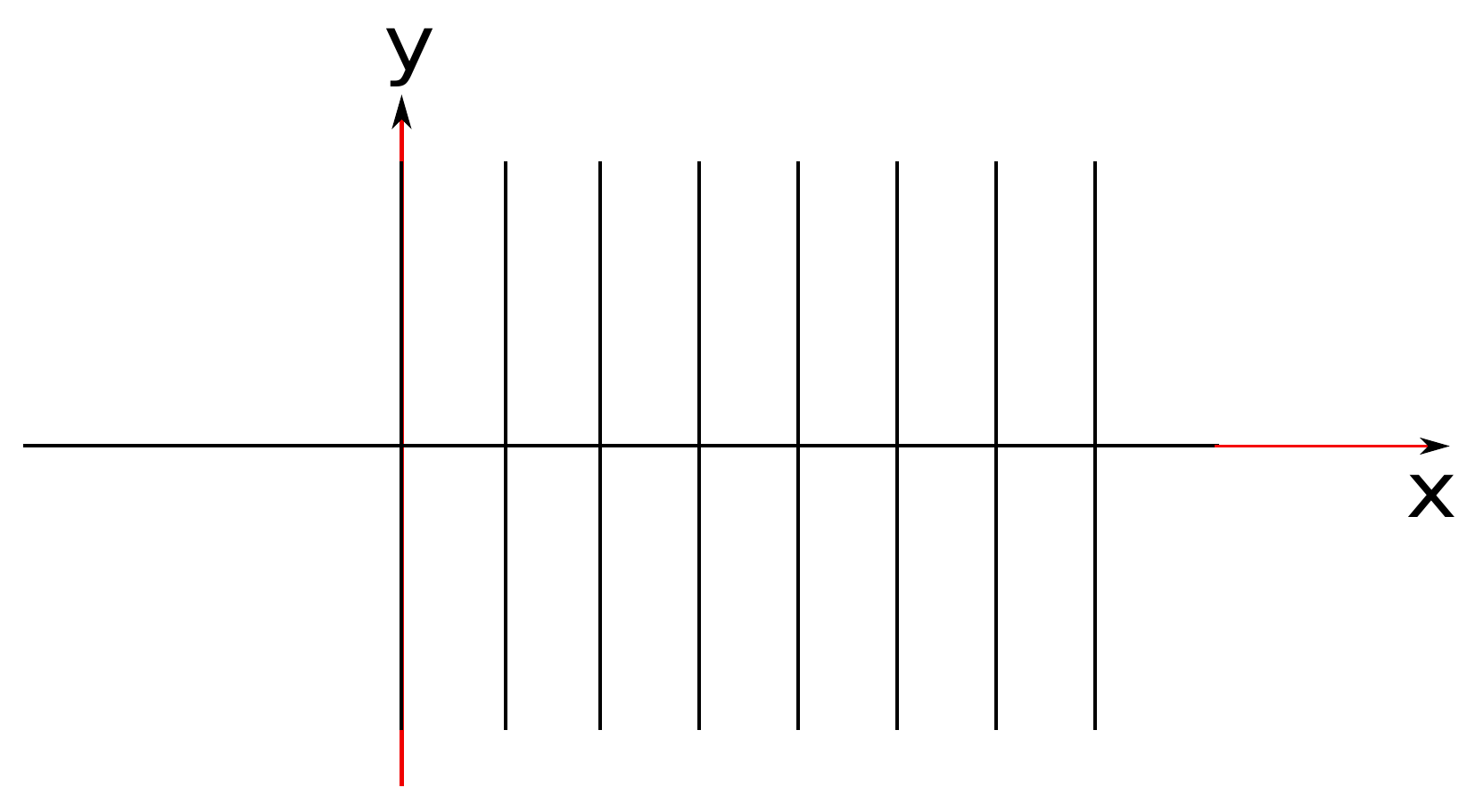}
\caption{The $2D$ comb model.}
\label{2DCombModel}
\includegraphics[scale=0.35]{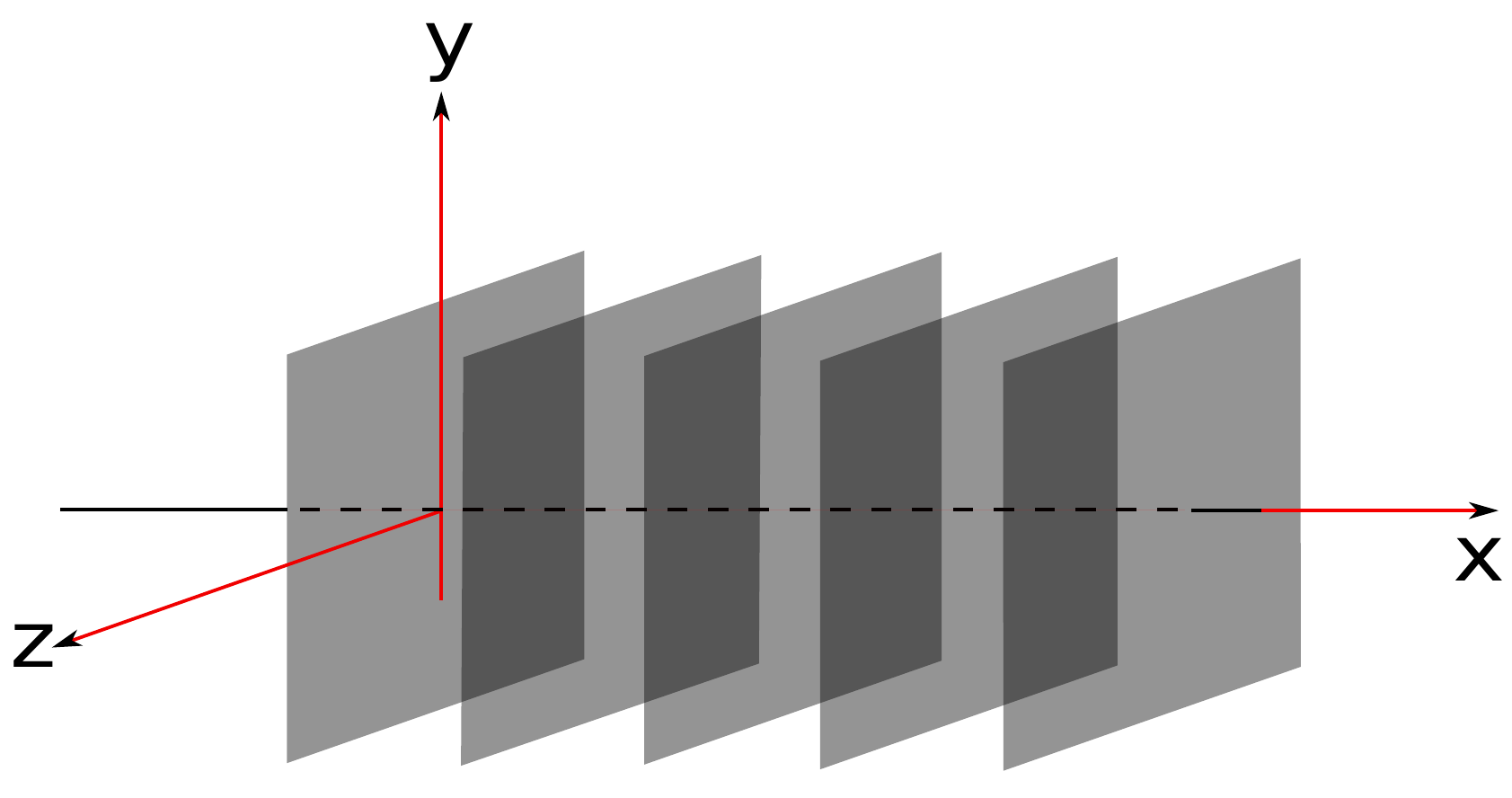}
\caption{The $3D$ model for comb with plane sidebranches with the $1D$ backbone (dubbed ``kebab'').}
\label{3DPlaneModel}
\end{figure}

Diffusion processes in complex systems may exhibit anomalous behavior \cite{Weiss1986, KlaS, MerS, Arkhincheev1991, Montroll1973, Montroll1984,
Gefen1983, Bouchaud1990, Sokolov}. Generally, by the term anomalous is meant that the diffusion front propagates slower or faster than the classical diffusion equation predicts.
Typically, when we consider the MSD of a randomly walking (or diffusing) particle we have $\left<x^2(t)\right>\propto t$ while in 
the anomalous diffusion one finds $\left<x^2(t)\right>\propto t^{\alpha}$, where $\alpha<1$ corresponds to subdiffusion, and 
$\alpha>1$ corresponds to super-diffusion.
The anomalous diffusion is observed in many fields such as: dusty plasma, polymer physics, financial systems, etc, see e.g. 
\cite{KlaS, Sokolov2012, GoreePRL, Zaburdaev, Mend1}. 
Comb structures are often considered as simple models of anomalous diffusion 
induced by geometric restrictions. In Fig.\ref{2DCombModel}, and Fig.\ref{3DPlaneModel} we show the classical comb 
structures that consist of a backbone decorated by identical arbitrary-shaped sidebranches depicted as lateral objects. The simplest 
two-dimensional ($2D$) comb was introduced by Weiss and Havlin \cite{Weiss1986} as a model for diffusion in the backbone of the percolation cluster, 
and was under continuous investigation since then \cite{Coniglio1981,Coniglio1982}.
Although the similarity between the comb and the percolation cluster is a bit superficial, the model itself is a very interesting and allows
for the discussion of its relation to the continuous time random walks \cite{KlaS}, aging phenomena \cite{MerS} etc. 
Anomalous transport on comb-like structures can also be formulated in terms of the Fokker-Plank equation \cite{Arkhincheev1991}. 

The further generalizations of the comb model onto more complex structures that consist of the backbone with sidebranches of different shapes
have led to very versatile general models of geometrically induced complex diffusions. These generalized models might be considered as the
limiting cases of different variants of transport in a tubes with dead ends and in tubes of varying cross-section \cite{Dagdug, Bere1,Vazquez, Bere2,
BerDugChan, Berezhkovsky, BerComb2}. 

The $2D$ comb model (Fig.\ref{2DCombModel}) with infinite sidebranches describes the subdiffusion along the backbone with 
$\left<x^2(t)\right>\propto t^{1/2}$ \cite{White1984, Weiss1986, Arkhincheev1991}. For finite sidebranches in the $2D$ comb model, 
subdiffusion is a transient process taking place at times shorter than some crossover time $t_0$ depending on the sidebranch lengths; 
for $t \gg t_0$ the transport along the backbone converges to normal diffusion \cite{Bouchaud1990}. Similarly, for the standard 
three-dimensional ($3D$) comb model (Fig.\ref{3DPlaneModel}) 
with finite sidebranches, the anomalous regime in the longitudinal diffusion is transient \cite{Forte2013,
IominMendez}.

In Ref. \cite{IominMendez} it was analytically shown that in the standard $3D$ cylindrical comb model with infinite sidebranches
the particles\textquotesingle{} spread along the one-dimensional ($1D$) backbone is ultra-slow: $\left<x^2(t)\right>\propto\ln t$. For this model, the propagator of the Fokker-Planck 
equation was found to be singular at points of junction of discs and $1D$ backbone, which leads to some mathematical 
difficulties in discussion of the Fokker-Plank equation in the Laplace space \cite{IominMendez}.  
The same ultraslow behavior was however obtained in \cite{Forte2013} based on scaling arguments for the return probability. Thus, \cite{Forte2013} 
gives an independent proof and explanation of the result. Ref. \cite{IominMendez} also states that if the motion in the side-branches is subdiffusive with 
$\left<x^2(t)\right>\propto t^{\alpha}$ for $0<\alpha<1$, the ultra-slow (logarithmic) diffusion changes to a 
considerably faster albeit still subdiffusive behavior $\left<x^2(t)\right>\propto t^{1-\alpha}\ln t$. However, the thought-provoking work \cite{IominMendez} failed 
to give a simple physical explanation of the obtained results. 

In the present paper we demonstrate how the diffusion behavior in the backbone depends on the assumptions on the junction
points between the sidebranches and the backbone. We note that previous works do not seem to pay the necessary attention to the assumptions about the junction points.  
We show how the peculiar speed-up of diffusion on the backbone, when the motion in the sidebranches
gets subdiffusive, arises within a very specific junction model. 
In our analysis, we first provide the random walk description of a diffusion in the $3D$ comb model with normal diffusion in the sidebranches, and with the anomalous 
diffusion thereon described by the CTRW. We discuss, under what assumptions the result of \cite{IominMendez} is reproduced, and what the alternative 
(and physically more plausible) assumption can be. We moreover discuss how various assumptions about the type of the junction may lead to various formulations 
of the Fokker-Planck equations that evidently possess different solutions.
\begin{figure*}
\centering
\includegraphics[scale=0.5]{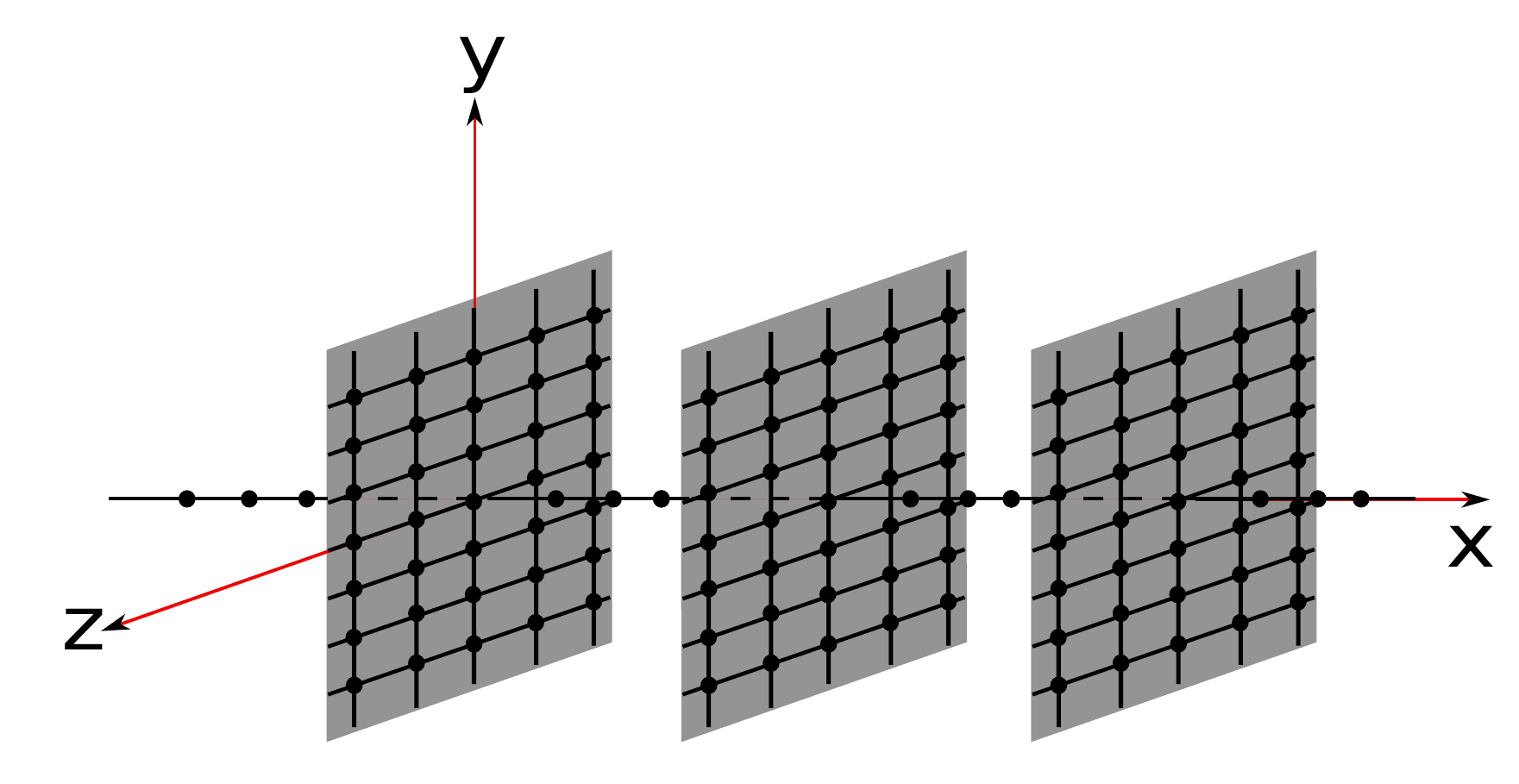}
\caption{$3D$ comb-lattice model}
\label{3DCombLattMod}
\end{figure*}

In addition, we analyze the case of the $3D$ cylindrical comb model as described by the Fokker-Planck equation \cite{IominMendez}. 
We discuss the nature of the analytical difficulties encountered in \cite{IominMendez}, and show that the full discussion of the solutions  
is possible in the Laplace space. We also study the regularization procedure that leads to the exact asymptotic solution, which, interestingly, coincides with the solution obtained within the random walk scheme.

\section{Random walks in comb lattices, and the junction models}
We consider the simplest random walk model in the $3D$ comb-lattice model (Fig.\ref{3DCombLattMod}) in which the backbone and the plane sidebranches of the ``kebab structure'' (called ``discs'' in what follows) are a $1D$ lattice and a $2D$ square lattice respectively. In our analysis, for the sake of simplicity, all lattice constants are
set to unity.

\subsection*{The model of a rejuvenating junction}
In comb models, the motion in the backbone is possible only when the particle, having made an excursion to the sidebranch, returned
to the backbone. In our first model, having arrived at a junction point, the particle is assigned a new waiting time, and the prehistory (i.e 
fact that the jump in the backbone took place during the waiting period in the sidebranch) is forgotten. This defines a true CTRW scheme for the 
corresponding lattice model. In this respect the junction point is not different from any other point. 
The diffusion process in the backbone can be approximated by the CTRW with waiting times related to
the first return times of the walker to its initial position (origin) in the sidebranches, i.e. to the junction points of the disc to the backbone. 
In this model the total time available for the diffusion in the backbone is proportional to the total number of visits of the origin. 
The model of junction that satisfies these assumptions will be called a rejuvenating junction. 

In the simplest case we can assume that being at the site at which the backbone is connected to the disc the walker performs the
jump to each of 6 directions with equal probability, the probability to leave the disc at return is 1/3, and the probability to stay is 2/3.
Let the waiting time for the step when being in the backbone be $\psi_b(t)$. This may differ from the waiting time in the 
disks. For example, the diffusion in the backbone might be normal, and in this case $\psi_b(t)$ possesses the finite first moment. 
Let us calculate $\phi(t)$, the probability distribution function (PDF) of the waiting time until the next step in a backbone. If this is known, 
the MSD along the backbone can be easily obtained in the Laplace domain (see Eq. (3.14) in \cite{KlaS}) :
\[
 \langle x^2(s) \rangle = \frac{\phi(s)}{s[1-\phi(s)]},
\]
where $\phi(s)$ is the Laplace transform of the first return time distribution $\phi(t)$ above.

Let us consider the particle starting at the backbone and entering the disk. Let $f(t)$ be the waiting time until it returns from the 
disk to the backbone, and ``decides'' to make a step thereon. Let $n$ be the number of excursions to the disk performed before the step in the
backbone is actually taken. If no excursion takes place this waiting time is given by $\psi_b(t)$, and such case has a probability $q$ (in our simple
lattice picture this will be 1/3).
With probability $p=1-q$ (2/3 in our lattice picture) an excursion takes place. Let the PDF of the excursion time be given by the function
$F(t)$. Then, the waiting time for the next step in a backbone
after excursion is given by $\psi_1(t) = \displaystyle\int_0^t F_1(t') \psi_b(t-t') dt'$. The excursion process is terminated after the first
excursion with probability $q$,
and with probability $p$ the next, second excursion takes place. If $n$ excursions take place before the step in the backbone is 
actually taken, the corresponding PDF is given by $\psi_n(t)$ being the $n$-fold convolution of $\psi_1$ with itself: $\psi_n(t) =
\psi_1(t)*...*\psi_1(t)$.
The probability of the $n$ excursions before the step in the backbone is $p^n$. Taking into account the convolution structure of all
terms, we get in the Laplace domain 
\[
 \phi(s) = q \psi_b(s) +  q \left(p F(s)\psi_b(s)\right) +   q p^2 F^2(s)\psi_b^2(s) + ... 
\]
this geometric series sums up to 
\begin{equation}
\begin{split}
\phi(s) &= q \psi_b(s) + q \sum_{k=1}^\infty \left(p F(s) \psi_b(s) \right)^k = \\ & q \psi_b(s) + q \frac{p F(s)\psi_b(s)}{1- p
F(s)\psi_b(s)}.
\end{split}
\end{equation}
This is correctly normalized: since $F(0) = 1$ and $\psi_b(0)=1$ in the Laplace domain one also has $\phi(0) = 1$ due to the fact that
$p+q=1$. 

Our next step will be to consider $F(t)$.
Let $F_n$ be the sequence of probabilities in a simple random walk in the $2D$ to return to the origin after $n$ steps. Let $\psi(t)$ be the
waiting time PDF for continuous-time
random walk in the disk. Since the return can only happen at the time of the step 
\[
F(t) = \sum _{n=0}^\infty F_n \psi_n (t) 
\]
where $\psi_n(t)$ is the $n$-fold convolution of $\psi(t)$ with itself, and has the Laplace representation $\psi_n(s) = \psi^n(s)$ (see p. 53
in \cite{KlaS}).
In the Laplace domain
\[
 F(s) = \sum _{n=0}^\infty F_n \psi^n(s).
\]
Introducing the generating function of the return probabilities, 
\[
 G(z) = \sum_{n=0}^\infty F_n z^n,
\]
 we see that
\[
 F(s) = G(\psi(s)). 
\]
The generating function of the return probabilities is connected with the generating function of the probability to be at the origin after $n$
steps $P(0,z)$ (Eq. (2.11) in \cite{KlaS}):
\[
 G(z) = 1 - \frac{1}{P(0,z)},
\]
where $P(0,z)$ is given by the Eq. of the Exercise 2.5 in \cite{KlaS}:
\[
 P(0,z) = \frac{1}{(2\pi)^2} \int_{-\pi}^\pi \int_{-\pi}^\pi \frac{1}{1-z \lambda(\mathbf{k})} d \mathbf{k}.
\]
On our lattice $\lambda(\mathbf{k}) = \displaystyle(\cos k_x + \cos k_y)/2$. We are interested in the behavior of the integral for $z$ close to
unity, where the integral diverges. The type of divergence can be seen when passing to polar coordinates close to $\mathbf{k} = 0$ when
$\lambda(\mathbf{k}) \approx 1 -k^2/4$
(and separating the non-singular part of the integral):
\begin{equation}
\begin{split}
 P(0,z) &\simeq \frac{1}{2\pi} \int_0^{k_{max}}  \frac{1}{1-z (1 - k^2/4) } k dk  + \\ & \{\mbox{non-singular part}\}
\end{split}
\end{equation}
The leading term diverging at $z \to 1$ is 
\[
 P(0,z) \simeq  - \frac{1}{\pi} \frac{\ln(1-z)}{z}.
\]
Substituting $z = \psi(s)$, and assuming $\psi(s) = 1 - (s\tau)^\alpha + ...$, where $\alpha$ is the exponent of the anomalous diffusion in a
plane
(the case $\alpha = 1$ corresponds to the normal diffusion), we get
for $s \to 0$
\[
 F(s) = 1 + \frac{\pi \psi(s)}{\ln[1 - \psi(s)]} \simeq 1 + \frac{\pi}{\alpha \ln s}.
\]
We note that this is an extremely weak $s$-dependence, which, for $s$ small enough will dominate over a whatever power law stemming from
$\psi_b(s) \simeq 1 - As^\alpha$.
The dominant behavior for small $s$ is thus
\[
 \phi(s) \simeq 1 + \frac{p}{1-p} \frac{\pi}{\alpha \ln s}. 
\]
Thus, the asymptotics of $\langle x^2(s) \rangle$ reads
\[
 \langle x^2(s) \rangle = \frac{1}{s[1-\phi(s)]} \simeq \frac{\alpha}{\pi} \frac{1-p}{p}\frac{\ln (1/s)}{s}.
\]
Inverting this expression by means of the Tauberian theorem (p. 45 in \cite{KlaS}) we get 
\begin{equation}
 \langle x^2(t) \rangle \simeq \frac{A \alpha}{\pi} \ln t,
 \label{MSD1}
\end{equation}
with $A=(1-p)/p$ independently of $\psi_b(t)$. The logarithmic dependence on $t$ for $\alpha = 1$ was obtained in \cite{Forte2013}, and the
whole expression
coincides with the result of the continuous model of \cite{IominMendez} (with $\alpha = 1$) if one associates $A$ with the diffusion
coefficient: $A = 1/2D$. It is important to note that the behavior of the mean squared displacement in the backbone asymptotically
does not depend at all on the distribution of waiting time $\psi_b(t)$ in the junction sites.

\subsection*{The model of non-rejuvenating junction}
\label{Junctions}

Our previous model was based on the assumption that the step in the spine \textit{interrupts} the waiting period in the sidebranch and resets the waiting
time anew when returning to the same or to the other sidebranch. It is a truly renewal model, which rejuvenates the motion in the sidebranch under return. 
Now, we assume that the motions in the disk and in the backbone
are absolutely independent, and the waiting time in the sidebranch is not reset by steps in the backbone. This assumption is tacitly done in
a whatever model allowing for variable separation, like in the model discussed in \cite{IominMendez}. 
This means that the waiting phase for the motion in the side structure is not interrupted by the jump in the backbone. 
The jumps in the backbone take place whenever one is in the backbone. In other words, we say that the junction is non-rejuvenating, if $(i)$
the waiting period in the sidebranches is not affected by the motion in the backbone, and $(ii)$ the motion in the backbone is only
possible when the position in the sidebranch corresponds to the junction point. 

In this junction model the total physical time available for the motion in the backbone 
is equal to the total time spent at the origin of the sidebranches, i.e. to 
\[
 T(t) = \int_0^t P(0,t')dt'.
\]
In the CTRW in $2D$ the asymptotics of $P(0,t)$ is given by 
\begin{equation}
\begin{split}
 P(0,s) &= \frac{1 - \psi(s)}{s} \tilde{P}(0,z=\psi(s)) \simeq  \\ & \frac{1}{\pi} \frac{1 - \psi(s)}{s} \ln \left(\frac{1}{1-\psi(s)} \right)
\end{split}
\end{equation}
(see Eq. (3.13) in \cite{KlaS}), with $\tilde{P}(0,z)$ being the generating function for the probability of being at the origin for a
simple random walk,
i.e.
\[
 P(0,s) \simeq \frac{\alpha}{\pi} s^{\alpha-1} \ln (1/s),
\]
which in the time domain translates into
\[
 P(0,t) \simeq \frac{\alpha}{\pi \Gamma(1-\alpha)} t^{-\alpha} \ln t.
\]
The corresponding integral $T(t)$ is then
\begin{equation}
\begin{split}
T(t) &= \frac{\alpha}{\pi \Gamma(1-\alpha)} \int_0^t (t')^{-\alpha} \ln t' = \\ & \frac{\alpha}{\pi \Gamma(1-\alpha)} t^{1-\alpha}
\left[\frac{\ln t}{1-\alpha} - \frac{1}{(1-\alpha)^2}\right].
\end{split}
\end{equation}
Asymptotically, for $t$ large, the second term in the square brackets can be neglected, and we get
\begin{equation}
\begin{split}
 T(t) &= \frac{\alpha}{\pi (1-\alpha) \Gamma(1-\alpha)} t^{1-\alpha} \ln t = \\ & \frac{\alpha}{\pi \Gamma(2-\alpha)} t^{1-\alpha} \ln t.
\end{split}
\end{equation}
If the diffusion in the backbone is normal, the MSD $\langle x^2(T) \rangle = D_0 T$ reads
\begin{equation} 
\langle x^2(t) \rangle =  \frac{\alpha D_0}{\pi \Gamma(2-\alpha)} t^{1-\alpha} \ln t.
\label{CTRW_MSD}
\end{equation}
After the appropriate redefinition of the constants this reproduces the result of \cite{IominMendez}. 
Note that $T(t)$ is the analog of the operational time for the diffusion in a backbone, and that this does not follow the CTRW scheme, 
i.e. is not a renewal process anymore.
Thus, the result we get shows that when the diffusion in the discs gets continuously slower, the diffusion in a backbone gets faster due to the fact that more 
time is spent in the backbone in total. 

If the diffusion in the backbone is also anomalous, with the exponent of anomalous diffusion $\beta$ we get $\langle x^2(T) \rangle \propto
T^\beta$,
and therefore 
\[
 \langle x^2(t) \rangle \propto t^{(1-\alpha)\beta} \ln^\beta t. 
\]
In this case the anomaly of diffusion in the backbone changes the overall diffusive behavior thereon. Note that for $\alpha = \beta$ we should get
\begin{equation}
 \langle x^2(t) \rangle \propto t^{(1-\alpha)\alpha} \ln^\alpha t. 
 \label{eq:alphabeta}
\end{equation}


\section{Dynamics in the modified infinite $3D$ cylindrical comb}
\label{SectDyn}
One of the main features of random walks in $3D$ comb lattices is that, due to discrete nature of these structures, the dynamics in them does not show singularities.
\begin{figure*}
\centering
\includegraphics[scale=0.5]{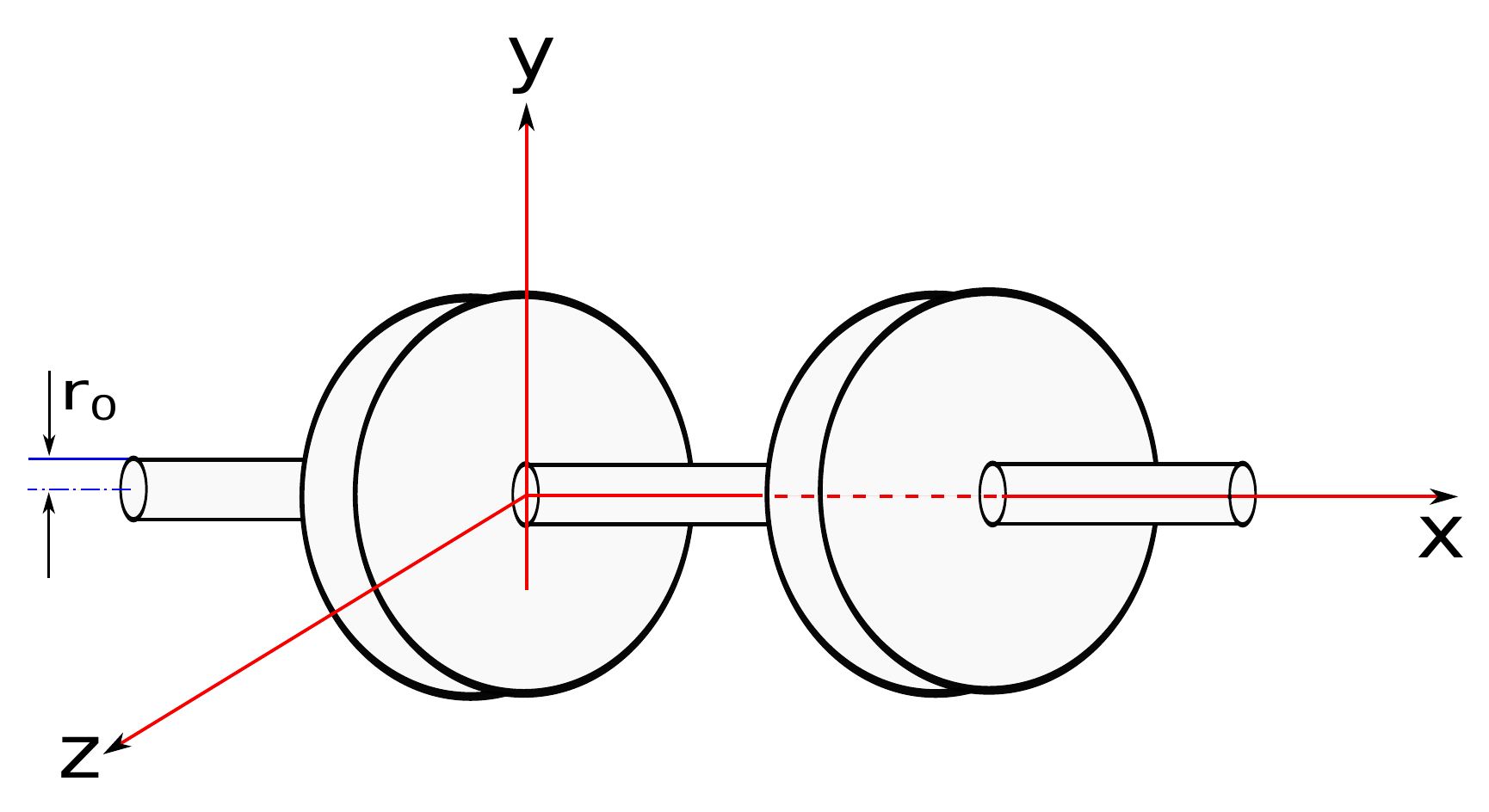}
\caption{The modified infinite $3D$ cylindrical comb model with the $2D$ cylindrical backbone of radius $r_0$: $y^2+z^2=r_0^2$}
\label{3DCombModelMod}
\end{figure*}
However, when we turn to a consideration of diffusion in continuos $3D$ comb models the possible singularity of the propagator for the
Fokker-Planck equation in the sidebranch does not allow us to correctly define the MSD \cite{IominMendez}. This singularity arises due to 
the assumption of the zero radius of the backbone (or of the particle). From the physical point of view the singularity arises due to the fact that a particle of zero size 
which has started at the origin of the sidebranch never hits this \textit{point} again, and therefore stays in a sidebranch indefinitely long. 
Assuming the finite radius of the spine tube (or of the particle), removes this singularity, and leads to the return to the corresponding area 
with probability one. On the other hand, the dependence on the radius is so weak (logarithmic) that it does not affect the final asymptotic behavior. 
Thus, for such comb models we need to apply the regularization techniques, assume the finite radius of the backbone tube, and show that this does
not influence the asymptotic behavior. As already stated, the various junction models imply different forms of the Fokker-Planck equation for the comb structure. 
To be specific, in this section, we focus on the Fokker-Planck equation for the case of non-rejuvenating junction as discussed in \cite{IominMendez}.
The detailed discussion of the possible differences between various formulations of the Fokker-Planck equation is given in the
Section \ref{Append}.

The random dynamics in the $3D$ comb structure is described by the 3D probability distribution function 
$P=P(x,y,z,t)$ of finding a particle at time $t$ at position $(y,z)$ in the $2D$ sidebranch that crosses the backbone 
along the $x$ axis. In this section, using the Fokker-Plank framework, we study the case
of a cylindrical comb that consists of a backbone of a small radius $r_0$ radius centered along the $x$ axis and decorated by an array of sidebranches of infinite radius,
see Fig.\ref{3DCombModelMod}. A modification of the backbone shape in the $3D$ cylindrical comb model serves as a regularization that exploits the
geometry of the model. This geometric regularization allows for obtaining a non-singular solution in sidebranches, and
consequently the correctly defined, finite MSD.

The diffusion equation in the dimensionless variables and parameters \cite{IominMendez}, in Cartesian coordinate system, reads
\begin{equation}
\partial_t P = D \hat{\Delta}_{xyz}P,
\label{from_paper}
\end{equation}\\
\noindent where $\hat{\Delta}_{xyz}\equiv D^{-1}\hat{\partial}_x^2 + \hat{\Delta}_{yz} = (1/2\pi
r_0)D^{-1}\delta\left(g(y,z)-r_0\right)\cdot\partial_x^2 + \Theta\left(g(y,z)-r_0\right)\times\left(\partial_{y}^2+\partial_{z}^2\right)$, with
$g(y,z)=\sqrt{y^2+z^2}$, and $r_0\in \mathbb{R}^{+}$, $(y,z)\in\mathbb{R}^1$, and $\Theta(u)$ is a Heaviside step function. We note that
$\displaystyle\frac{1}{2\pi r_0}\displaystyle\int \int\delta\left(g(y,z)-r_0\right) dydz=1$. In accordance with the geometry of the modified
comb model, the shift in $x$ direction is possible only when $(y,z)\in\{g(y,z)-r_0=0\}$, i.e. on the surface of cylinder with radius $r_0$
along the $x$ axis, and the shift in $y$-$z$ plane is possible in any direction except $(y,z)\in\{g(y,z)-r_0<0\}$.The initial condition is
\begin{equation}
P_0\equiv P(x,y,z,t=0) =\displaystyle\frac{1}{2\pi r_0}\delta(x)\cdot\delta\left(g(y,z)-r_0\right).
\label{initial_cond}
\end{equation}
The formal solution of the Eq. (\ref{from_paper}) in a convolution form is
\begin{equation}
P(x,y,z,t) = \int_0^tG(y,z,t-t^{'})F(x,t^{'})dt,
\label{convolution} 
\end{equation}
where $G(y,z,t)$ is the propagator for two dimensional diffusion in the sidebranches (discs), and $F(x,t)$ relates to the solution along the
backbone. 
The solution in a form (\ref{convolution}) corresponds to the case when the motion in sidebranches is not affected by the motion in the
backbone. In other words, it implements the first condition $(i)$ of the non-rejuvenating junction.
Performing the Laplace transform of Eq. (\ref{from_paper}) one obtains
\begin{equation}
s\widetilde{P}=\hat{\partial}_x^2 \widetilde{P} + D\hat{\Delta}_{yz}\widetilde{P} + P_0.
\label{Laplace_domain}
\end{equation}
The Eq. (\ref{convolution}) in the Laplace domain reads as
\begin{equation}
\widetilde{P}(x,y,z,s) = \widetilde{G}(y,z,s) \widetilde{F}(x,s)\equiv \widetilde{G}(r,s) \widetilde{F}(x,s).
\label{convol_in_Laplace}
\end{equation}
In the last equality we used the cylindrical symmetry of the model. From Eq. (\ref{Laplace_domain}) and Eq. (\ref{convol_in_Laplace}) 
we obtain the equation for the diffusion in a disc, which, in cylindrical coordinates, reads as
\begin{equation}
s\widetilde{G} = \displaystyle D\frac{1}{r}\displaystyle\frac{\partial}{\partial r}\left(r\displaystyle\frac{\partial \widetilde{G}}{\partial
r}\right),
\label{Diff_disc} 
\end{equation}
with $r\in[r_0,\infty)$. In the junction of a disc and backbone we set
\begin{equation}
\widetilde{G}|_{r=r_0} = \widetilde{G}_0(x,s)\equiv\widetilde{G}_0,
\label{boundary}
\end{equation}
This expression means that the motion in the backbone is only possible when the position in the sidebranch corresponds to the junction points.
It implements the second condition $(ii)$ of the non-rejuvenating junction.
The solution of Eq. (\ref{Diff_disc}) which satisfied the boundary condition at $r\rightarrow \infty$ may be expressed through the modified
Bessel functions of the second kind:
\begin{equation}
\widetilde{G}(r,s) = \widetilde{G}_0 \frac{K_0(r\sqrt{s/D})}{K_0(r_0\sqrt{s/D})}.
\label{DiscSolution}
\end{equation}
It is clearly seen from (\ref{DiscSolution}) that in the case of the $3D$ cylindrical comb model with $1D$ sidebranch, this
solution is singular, since the sidebranch radius $r_0 = 0$.\\
To define the MSD in the $x$ direction, one needs to find a reduced distribution $P_1(x,t)$ in the Laplace space. From
Eq. (\ref{convol_in_Laplace}) and Eq. (\ref{DiscSolution}) we have
\begin{eqnarray}
\begin{split}
\widetilde{P_1}(x,s) & \equiv \int_{r_0}^{+\infty}2\pi r dr\widetilde{P}(x,r,s) =\\& 2\pi\widetilde{F}(x,s) \int_{r_0}^{+\infty}
rdr\widetilde{G}(r,s) =\\& \frac{2\pi\widetilde{F}(x,s)\widetilde{G}_0}{K_0(r_0\sqrt{s/D})}\int_{r_0}^{+\infty} rdr K_0(r\sqrt{s/D}) = \\& 2\pi
\sqrt{D}\widetilde{G}_0 r_0\cdot\frac{\widetilde{F}(x,s) K_1(r_0\sqrt{s/D})}{\sqrt{s} K_0(r_0\sqrt{s/D})},
\label{reducedDF}
\end{split}
\end{eqnarray}
since \cite{GradshteynRyzhik}: $\displaystyle\int_{r_0}^{\infty} uK_0(au) du = \frac{r_0}{a}\cdot K_1(a r_0)$. Thus,
\begin{equation}
\begin{split}
\widetilde{G}_0(x,s)\widetilde{F}(x,s) =\qquad \qquad \qquad \qquad \qquad \qquad \\ \frac{s\widetilde{P_1}(x,s)}{2\pi D} \cdot
\frac{K_0(r_0\sqrt{s/D})}{r_0\sqrt{s/D} K_1(r_0\sqrt{s/D})} =\\\frac{s\widetilde{P_1}(x,s)}{2\pi D} \cdot \hat{K}(r_0\sqrt{s/D}),
\end{split}
\label{FKP}
\end{equation}
here we denote $\hat{K}(r_0\sqrt{s/D}) \equiv \displaystyle\frac{K_0(r_0\sqrt{s/D})}{r_0\sqrt{s/D} K_1(r_0\sqrt{s/D})}$.
Now, taking into account the Eq. (\ref{DiscSolution}) we integrate Eq. (\ref{Laplace_domain}) over ($y,z$) - domain and after the transition to
polar coordinates $(r, \varphi)$ one gets
\begin{equation}
\begin{split}
s\widetilde{P}_1(x,s)&=\displaystyle\frac{1}{2\pi r_0}\partial_x^2\int_{0}^{\infty}2\pi r dr \delta(r-r_0)\widetilde{F}(x,s)\times
\\&\frac{\widetilde{G}_0 K_0(r\sqrt{s/D})}{K_0(r_0\sqrt{s/D})} + \delta(x).
\label{int_Laplace_domain}
\end{split}
\end{equation}
The integration in Eq. (\ref{int_Laplace_domain}) gives
\begin{equation}
s\widetilde{P}_1(x,s)=\partial_x^2\left[\widetilde{G}_0(x,s)\widetilde{F}(x,s)\right] + \delta(x).
\label{After_int_Laplace_domain}
\end{equation}
Substituting the Eq. (\ref{FKP}) into Eq. (\ref{After_int_Laplace_domain}) and using (\ref{boundary}) we have
\begin{equation}
s\widetilde{P}_1(x,s)= \frac{1}{2\pi D}s\hat{K}(r_0\sqrt{s/D})\cdot\partial_x^2 \widetilde{P}_1(x,s) + \delta(x).
\label{Subst_int_Laplace_domain}
\end{equation}
The Fourier transform of Eq. (\ref{Subst_int_Laplace_domain}), leads to
\begin{equation}
\overline{\widetilde{P}}_1(k,s) = \frac{2\pi D}{s[2\pi D + k^2\hat{K}(r_0\sqrt{s/D})]}.
\label{Fourier_domain}
\end{equation}
The Eq. (\ref{Fourier_domain}) yields the MSD in the Laplace domain in the form 
\begin{equation}
\left<\widetilde{x^2}(s)\right> = \left[ -\left. \frac{d^2}{dk^2}\overline{\widetilde{P}}_1(k,s)\right\vert_{k=0} \right].
\end{equation}
From Eq. (\ref{Fourier_domain}) we have 
\begin{equation}
\left<\widetilde{x^2}(s)\right> = \left[ \frac{1}{2\pi D}\cdot\frac{2\hat{K}(r_0\sqrt{s/D})}{s}\right].
\label{MSD_def}
\end{equation}
Here, we are interested in the long time dynamics i.e. when $s\rightarrow 0$. In this case, $K_0(x)/K_1(x) \sim x\ln\displaystyle\frac{x}{2}$,
and $2\hat{K}(r_0\sqrt{s/D}) \sim \ln s + \ln(r_0^2/4D)$. 
\begin{eqnarray}
\begin{split}
\left<\widetilde{x^2}(s)\right>  &=  \left[ \frac{1}{2\pi D}\cdot\frac{1}{s}\left(\ln s + \ln(r_0^2/4D)\right)\right] =\\&
\displaystyle\frac{1}{2\pi D} \left[s^{-1}\ln s\right] + \frac{1}{2\pi D} \ln\left( \frac{r_0^2}{4D} \right)\left[\frac{1}{s}\right],
\label{MSD_calc}
\end{split}
\end{eqnarray}
and, for $s \to 0$, is dominated by the first term:
\begin{equation}
 \langle \widetilde{x^2}(s) \rangle \simeq \frac{1}{2\pi D} \frac{\ln s}{s}.
 \label{eq:LapUS}
\end{equation}
Going back to the time domain \cite{Bateman1954, Jahnke1960} we get 
\begin{equation}
\left<x^2(t)\right> \sim \frac{1}{2\pi D}\ln t,
\label{MSDR}
\end{equation}
and ultra-slow diffusion takes place, as expected \cite{Forte2013}. It is easy to see that the expression (\ref{MSDR}) holds also true for
$r_0\rightarrow 0$. So, we have illustrated, that the cylindrical $3D$ comb model with $1D$ cylindrical backbone can be considered as a limit
case of the modified cylindrical $3D$ comb model, when $r_0\rightarrow 0$. We also show that the MSD can be correctly defined for the
cylindrical $3D$ model. Thus, the correct physical result is obtained when the regulator $r_0$ vanishes.

\subsection*{Anomalous diffusion in sidebranches}
\label{AnomalousDiscs}

Now, we consider the anomalous diffusion in discs by to the generalizating Eq. (\ref{Diff_disc}) in the Laplace domain: 
\begin{equation}
s\widetilde{G} = \displaystyle s^{1-\alpha}D\frac{1}{r}\displaystyle\frac{\partial}{\partial r}\left(r\displaystyle\frac{\partial
\widetilde{G}}{\partial r}\right),
\label{Diff_disc_frac} 
\end{equation}
with the boundary condition (\ref{boundary}). Consequently, it leads to the solution
\begin{equation}
\widetilde{G}(r,s) = \widetilde{G}_0 \frac{K_0(r\sqrt{s^{\alpha}/D})}{K_0(r_0\sqrt{s^{\alpha}/D})}.
\label{DiscSolution_frac}
\end{equation}
Then, we have
\begin{equation}
\begin{split}
\widetilde{G}_0(x,s)\widetilde{F}(x,s) = \qquad \qquad \qquad \qquad \qquad \\ \frac{s^{\alpha}\widetilde{P_1}(x,s)}{2\pi D} \cdot
\frac{K_0(r_0\sqrt{s^{\alpha}/D})}{r_0\sqrt{s^{\alpha}/D} K_1(r_0\sqrt{s^{\alpha}/D})} =\\ \frac{s^{\alpha}\widetilde{P_1}(x,s)}{2\pi D} \cdot
\hat{K}(r_0\sqrt{s^{\alpha}/D}), \qquad \qquad \qquad
\end{split}
\label{FKP_frac}
\end{equation}
and 
\begin{equation}
\begin{split}
s\widetilde{P}_1(x,s) = \quad \qquad \qquad \qquad \qquad \qquad \qquad \\ \frac{1}{2\pi
D}s^{\alpha}\hat{K}(r_0\sqrt{s^{\alpha}/D})\cdot\partial_x^2 \widetilde{P}_1(x,s) + \delta(x).
\label{Subst_int_Laplace_domain_frac}
\end{split}
\end{equation}
Further, the Fourier transform of Eq. (\ref{Subst_int_Laplace_domain_frac}), leads to
\begin{equation}
\overline{\widetilde{P}}_1(k,s) = \frac{2\pi D}{s[2\pi D + k^2s^{\alpha-1}\hat{K}(r_0\sqrt{s^{\alpha}/D})]}.
\label{Fourier_domain_frac}
\end{equation} 
Finally, in the limit $s\rightarrow 0$, for the MSD, we have 
\begin{equation}
\begin{split}
\left<x^2(t)\right> &\sim \frac{1}{2\pi D}\frac{\alpha}{\Gamma(2-\alpha)}\cdot t^{1-\alpha}\ln t =\\& 
\frac{1}{2\pi \overline{D}}\cdot t^{1-\alpha}\ln t, 
\label{Fractional_finalMSD_frac}
\end{split}
\end{equation}
where $\overline D\equiv (D/\alpha)\Gamma(2-\alpha)$ is a generalized transport coefficient. The sub-diffusion with the transport exponent
$1-\alpha$ is dominant. The ultra-slow diffusion $\sim\ln t$ takes place only for $\alpha=1$ that can be realized as the result of normal
diffusion in the infinite sidebranched discs. As we can see, the result (\ref{Fractional_finalMSD_frac}) also holds for the case with $r_0=0$.

We note that the MSD (\ref{Fractional_finalMSD_frac}) in the regularized comb model exactly coincides with the MSD (\ref{CTRW_MSD}) obtained
for the random walk model under the same assumption on the nature of the junction.  As a result, we observe that the slow-down of 
diffusion in the disc leads to speeding-up of the diffusion along the backbone, and to the corresponding change in the rate of growth of the MSD
(\ref{Fractional_finalMSD_frac}). This conclusion is in a correspondence with results obtained in \cite{Berezhkovsky}.

\section{Generalized Fokker-Planck approach for the model of rejuvenating junction.}
\label{Append}
The difference between the junction models is clearly seen when we consider the generalized master equations that govern the corresponding
motion.

Our model of rejuvenating junction is the CTRW model in a comb lattice. The approach to such models with position-dependent
waiting time distributions was discussed in 
\cite{Chechkin} giving generalization of
fractional diffusion equations to inhomogeneous media. Let $\hat{L}$ be the Laplace matrix of our discrete lattice structure, 
i.e. of the comb. Then the evolution equation for the vector of probabilities $p_i$ to find a particle in the site $i$ of the structure
$\mathbf{p} = \left(p_1,...,p_N\right)$
is given by
\[
 \frac{d}{d t} \mathbf{p}(t)  = \hat{L}\int_0^t \mbox{\boldmath$\Phi$}(t-t') \circ \mathbf{p}( t') dt'
\]
where $\mbox{\boldmath$\Phi$}$ is the vector of the memory kernels given by their Laplace transforms:
\[
 \Phi_i(s) = \frac{s \psi_i(s)}{1 - \psi_i(s)}
\]
with $\psi_i(t)$ being the waiting time PDF in the site $i$ and $\circ$ denotes the Hadamard (entry-wise) product of two vectors, being a
vector of the same dimension
with elements $\Phi_i(t-t')p_i(t')$. In a comb with the different waiting times in the disks and in the backbone, there are only two different
$\Phi$:
$\Phi_d$ in the disk and $\Phi_s$ in the backbone. When passing to the continuous representation 
we get 
the Fokker-Planck equation with the position-dependent kernel: 
\begin{equation}
\begin{split}
\frac{\partial}{\partial t} P &= \left[ \delta(y)\delta(z) \frac{\partial^2}{\partial x^2} + D \left(\frac{\partial^2}{\partial y^2} +
\frac{\partial^2}{\partial z^2} \right) \right]\times \\ & \int_0^t \Phi(x,y,z,t-t') P(x,y,z, t') dt',
\end{split}
\end{equation}
where $\Phi$ depends on all three coordinates. Note that the spacial derivatives act on the kernel as well, and that the variables in this 
equation in general do not separate. 

If all waiting time distributions, in the junction sites and in the sites of the sidebranches, are the same, the memory kernel is position-independent: $\Phi(x,y,z,t) = \Phi(t)$. In this case
the memory operator commutes with the spatial derivatives, and the equation can be effectively rewritten  as
\begin{eqnarray}
\frac{\partial}{\partial t} P = \int_0^t dt' \Phi(t-t') \left[ \delta(y)\delta(z) \frac{\partial^2}{\partial x^2} + \right. \nonumber \\ 
\left. D \left(\frac{\partial^2}{\partial y^2} + \frac{\partial^2}{\partial z^2} \right) \right] P(x,y,z, t'),
\end{eqnarray}
but, due to the independence of the asymptotical behavior on the waiting times in the backbone discussed before, shows the same asymptotic solution.
This equation can be rewritten in the form
\begin{equation}
 \frac{\partial}{\partial t} (x,y,z, t) = \int_0^t dt'  \Phi(t-t') \hat{C} P(x,y,z, t'),
 \label{eqn:P}
\end{equation}
with the differential operator
\[
\hat{C} = \left[ \delta(y)\delta(z) \frac{\partial^2}{\partial x^2} + \right. \\ \left. D
\left(\frac{\partial^2}{\partial y^2} + \frac{\partial^2}{\partial z^2} \right) \right]
\]
describing the diffusion of the same comb structure for the case when the diffusion in the sidebranches and in the spine is normal. 
The solution of such Eq. (\ref{eqn:P}) is connected with the solution of the equation
\[
 \frac{\partial}{\partial t} F(x,y,z, t) = \hat{C} F(x,y,z, t)
\]
for the same initial condition by the integral transformation
\[
 P(x,y,z,t) = \int_0^\infty F(x,y,z,\tau) T(\tau,t) d \tau
\]
where $T(\tau,t)$ is the probability density of the number of steps done up to the time $t$ \cite{KlaS,Sokolov}, which is 
immediately connected to the memory kernel of Eq. (\ref{eqn:P}).
This relation looks the simplest in the Laplace domain
\begin{equation}
 \widetilde{P}(x,y,z,s) = \frac{1}{s M(s)} \widetilde{F}\left(x,y,z \frac{1}{M(s)} \right)
 \label{Lap1}
\end{equation}
with $\widetilde{P}(x,y,z,s)$ and $\widetilde{F}(x,y,z,s)$ being the Laplace transforms of $P(x,y,z,t)$ and $F(x,y,z,t)$ in their temporal variable.   
The function $M(s)$ is equal to $\Phi(s)/s$ and is connected with the Laplace transform $\psi(s)$ of the waiting time probability density by a relation
\begin{equation}
 M(s) = \frac{\psi(s)}{1-\psi(s)}.
 \label{eq:M}
\end{equation}
The function $M(t)$, the inverse Laplace transform of $M(s)$, has the physical meaning of the time-dependent rate of steps.
For $\psi(s) \simeq 1-s^\alpha$ it is given by $M(s) \simeq s^{-\alpha}$ for $s \to 0$. 

Note that the MSD in the backbone is a linear functional of $P(x,y,z,s)$ involving only integration over the spatial variables,
\[
  \langle x^2(t) \rangle = \int x^2 P(x,y,z,t) dx dy dz,
\]
and therefore the same relation holds between the MSD in the backbone of our structure and of a comb where diffusion is normal:
\[
 \langle \widetilde{x^2}(t) \rangle = \int_0^\infty \langle \widetilde{x^2}_s(\tau)\rangle T(\tau,t) d\tau
\]
(here $\langle x^2_s(\tau)\rangle$ is the MSD along the $x$-axis in the simple random walk (simple diffusion) in the same structure) or, in the Laplace domain
\[
 \langle \widetilde{x^2}(s) \rangle = \frac{1}{s M(s)}  \left\langle \widetilde{x^2}_s\left(\frac{1}{M(s)}\right) \right\rangle.
\]
The expression for $\langle \widetilde{x^2}_s(s)\rangle$ is given by Eq. (\ref{eq:LapUS}), and therefore we get:
\[
  \langle \widetilde{x^2}(s) \rangle \simeq \frac{1}{sM(s)} \frac{1}{2 \pi D} M(s) \ln \left(\frac{1}{M(s)}\right) = \frac{\alpha}{2 \pi D} \frac{\ln s}{s}.
\]
The inverse Laplace transform can be performed by using a Tauberian theorem and gives
\[
  \langle x^2(t) \rangle = \frac{\alpha}{2\pi D} \ln t,
\]
which reproduces our lattice result. 

\section{Conclusions}

In this work, we have investigated the effect of the junction model on the diffusion behavior in the comb structure
in the case when the diffusion in comb\textquotesingle s sidebranches is anomalous and is described by the CTRW. We have shown, that
the MSD in the backbone of the comb behaves differently depending on whether the waiting time periods
in the sidebranches are reset after the step in the backbone is done (a rejuvenating junction model), or not (a non-rejuvenating
junction). In the rejuvenating case the subdiffusion in the sidebranches only changes the prefactior in the ultra-slow (logarithmic) 
diffusion along the backbone, while in the non-rejuvenating case the ultraslow, logarithmic subdiffusion is changed to a 
much faster power-law subdiffusion (with a logarithmic correction). Moreover, in the first case the result does not change if the 
diffusion in the backbone is itself anomalous, while in the second case it does. 

Two of the considered cases were investigated within the Fokker-Planck approach. The first case is the one considered in
\cite{IominMendez}, where the appropriate regularization procedure allowed us to avoid some difficulties encountered in
the original work, and the another one corresponds to a special case of the rejuvenating junctions, where, again, the CTRW 
solution is reproduced. 

An open question stays how the anomalous diffusion in a backbone can be incorporated into the Fokker-Planck description of
a comb with non-rejuvenating junctions, since the corresponding process is not of a type of the processes discussed above. 



\section*{References}

\begin{thebibliography}{99}

\bibitem{Weiss1986}
Weiss G. H., Havlin S. 
Some properties of a random walk on a comb structure. 
Physica A 1986; 134:474.

\bibitem{KlaS} 
Klafter J., Sokolov I.M. 
First steps in Random Walks. 
Oxford: Oxford University Press; 2011.

\bibitem{MerS} 
Meroz Y., Sokolov I.M. 
Phys. Rev. Lett. 2011; 107: 260601.

\bibitem{Arkhincheev1991}
Arkhincheev V. E., Baskin E. M. 
Anomalous diffusion and drift in a comb model of percolation clusters.
Sov. Phys. JETP 1991; 73:161.

\bibitem{Montroll1973}
Montroll E. W., Scher H. 
Random walks on lattices. J. Stat. Phys. 1973; 9:101

\bibitem{Montroll1984}
Montroll E. W., Shlesinger M. F. 
The wonderful world of random walks. 
In: Lebowitz J., Montroll E. W., editors. Studies in statistical mechanics, vol. 11. Amsterdam: Noth-Holland; 1984.

\bibitem{Gefen1983}
Gefen Y., Aharony A., Alexander S. 
Anomalous diffusion on percolating clusters. 
Phys Rev Lett 1983; 50:77.

\bibitem{Bouchaud1990}
Bouchaud J.P., Georges A. Anomalous diffusion in disordered media: statistical mechanisms, models and physical applications. 
Phys. Rep. 1990; 195:127.


\bibitem{Sokolov} 
Sokolov I.M. 
Solutions of a class of non-Markovian Fokker-Planck equations. 
Phys. Rev. E 2002; 66:041101.


\bibitem{Sokolov2012}
Sokolov I. M. 
Models of anomalous diffusion in crowded environments. 
Soft Matter 2012; 8:9043.

\bibitem{GoreePRL}
B. Liu, J. Goree
Superdiffusion and Non-Gaussian Statistics in a Driven-Dissipative 2D Dusty Plasma.
Phys. Rev. Lett. 2008; 100:055003

\bibitem{Zaburdaev}
V.Yu. Zaburdaev, K.V. Chukbar
Enhanced superdiffusion and finite velocity of Levy flights.
JETP 2002; 94:252

\bibitem{Mend1} 
M\'endez V., Iomin A.  
Comb-like models for transport along spiny dendrites. 
Chaos, Solitons \& Fractals 2013; 53:46.


\bibitem{Coniglio1981}
Coniglio A. 
Thermal Phase Transition of the Dilute $s$-State Potts and $n$-Vector Models at the Percolation Threshold. 
Phys. Rev. Lett. 1981; 46:250

\bibitem{Coniglio1982}
Coniglio A. 
Cluster structure near the percolation threshold. 
J. Phys. A: Math. Gen. 1982; 15:3829 


\bibitem{Dagdug} 
Dagdug L., Berezhkovskii A.M., Makhnovskii Yu. A. et al. 
Transient diffusion in a tube with dead ends. 
J. Chem. Phys. 2007; 127:224712.

\bibitem{Bere1}
Berezhkovskii A.M., Dagdug L. 
Analytical treatment of biased diffusion in tubes with periodic dead ends. 
J. Chem. Phys. 2011; 134:124109.

\bibitem{Vazquez} 
Vazquez M.-V., Berezhkovskii  A.M., Dagdug L. 
Diffusion in linear porous media with periodic entropy barriers: A tube formed by contacting spheres. 
J. Chem. Phys. 2008; 129:046101.

\bibitem{Bere2}
Pineda I., Vazquez M.-V., Berezhkovskii A.M. et al. 
Diffusion in periodic two-dimensional channels formed by overlapping circles: Comparison of analytical and numerical results.  
J. Chem. Phys. 2011; 135:224101.

\bibitem{BerDugChan} 
Verdel R., Dagdug L., Berezhkovskii A.M. et al.  
Unbiased diffusion in two-dimensional channels with corrugated walls. 
J. Chem. Phys. 2016; 144:084106.

\bibitem{Berezhkovsky}
Berezhkovskii A. M., Dagdug L, Bezrukov S. M. 
From normal to anomalous diffusion in comb-like structures in three dimensions.
J Chem Phys. 2014; 141(5):054907.

\bibitem{BerComb2} 
Berezhkovskii A.M., Dagdug L., Bezrukov S.M. 
Biased diffusion in three-dimensional comb-like structures. 
J. Chem. Phys. 2015; 142:134101.

\bibitem{White1984}
White S. R., Barma M. 
Field-induced drift and trapping in percolation networks. 
J. Phys. A 1984; 17:2995.

\bibitem{Forte2013}
Forte G., Burioni R., Cecconi F., Vulpiani A. 
Anomalous diffusion and response in branched systems: a simple analysis. 
J. Phys. Condens. Matter 2013; 25:465106.


\bibitem{IominMendez}
Iomin A., M\'endez V.
Does ultra-slow diffusion survive in a three dimensional cylindrical comb?
Chaos Solitons and Fractals 2016; 82:142.

\bibitem{GradshteynRyzhik}
Gradshteyn I. S., Ryzhik I. M. 
Table of integrals, series, and products. 
Amsterdam: Elsevier Academic Press; 2007.


\bibitem{Bateman1954}
Bateman H, Erd\'elyi A. 
Tables of integral transforms, 1. 
New York: McGraw-Hill; 1954.

\bibitem{Jahnke1960}
Jahnke E, Emde F, L{\"o}sch F. 
Tables of higher functions. 
New York: McGraw-Hill; 1960.


\bibitem{Chechkin} 
Chechkin A.V., Gorenflo R., Sokolov I.M.
Fractional diffusion in inhomogeneous media. 
J. Phys. A: Math. Gen. 2005; 38(42):L679.
 

\end{thebibliography}
\end{document}